\documentclass[aps,prb,reprint,twocolumn]{revtex4-2}
\usepackage[dvipdfmx]{graphicx}
\usepackage{graphicx,latexsym}
\usepackage[dvipdfmx]{color}
\usepackage[dvipsnames]{xcolor}
\definecolor{DarkGreen}{rgb}{0.0, 0.5, 0.0}
\usepackage[colorlinks,citecolor=blue,linkcolor=blue,urlcolor=blue,linktocpage]{hyperref}
\usepackage{textcomp}
\usepackage{bm}
\usepackage{amsmath,amssymb,amsthm,amsfonts}
\usepackage{comment}
\usepackage{grffile}
\usepackage{cancel}
\usepackage{algpseudocode}
\usepackage{lipsum}
\usepackage{physics}
\usepackage{mathtools}
\newtheorem{lemma}{Lemma}

\newcommand{\vac}{\ket{\mathrm{vac}}}
\renewcommand{\ket}[1]{| #1 \rangle}
\renewcommand{\bra}[1]{\langle #1 |}

\begin{document}
%
%
\title{Construction of asymptotic quantum many-body scar states in the SU($N$) Hubbard model}
%
\author{Daiki Hashimoto}
\email{1225554@ed.tus.ac.jp}
\author{Masaya Kunimi}
\email{kunimi@rs.tus.ac.jp}
\author{Tetsuro Nikuni}
\affiliation{Department of Physics, Tokyo University of Science, 1-3 Kagurazaka, 
Tokyo 162-8601,  Japan}
%
%
%
\date{\today}
%
\begin{abstract}
    We construct asymptotic quantum many-body scars (AQMBS) in one-dimensional SU($N$) Hubbard chains ($N\geq 3$) by embedding the scar subspace into an auxiliary Hilbert subspace $\mathcal{H}_P$ and identifying a parent Hamiltonian within it, together with a corresponding extension of the restricted spectrum-generating algebra to the multiladder case.  Unlike previous applications of the parent-Hamiltonian scheme, we show that the parent Hamiltonian becomes the SU($N$) ferromagnetic Heisenberg model rather than the spin-1/2 case, so that its gapless magnons realize explicit AQMBS of the original model. Working in the doublon-holon subspace, we derive this mapping, obtain the one-magnon dispersion for periodic and open boundaries, and prove (i) orthogonality to the scar states, (ii) vanishing energy variance in the thermodynamic limit, and (iii) subvolume entanglement entropy with rigorous MPS/MPO bounds. Our results broaden the parent-Hamiltonian family for AQMBS beyond spin-1/2 and provide analytic, low-entanglement excitations in SU($N$)-symmetric systems.    
\end{abstract}
\maketitle
\section{Introduction}\label{sec:Introduction}
Recently, thermalization in isolated quantum many-body systems has attracted much attention. The eigenstate thermalization hypothesis~\cite{Deutsch1991,Srednicki1994,Rigol2008} provides a widely applicable paradigm for thermalization in nonintegrable systems. However, a certain class of nonintegrable systems, referred to as nonergodic systems, do not exhibit thermalization after long-time evolution. The typical examples include many-body localization~\cite{Gornyi2005Nov,Basko2007Aug,Nandkishore2015,abanin2019colloquium}, Hilbert space fragmentation  \cite{sala2020,Khemani2020,Moudgalya2022}, and quantum many-body scars (QMBS)~\cite{Bernien2017,Turner2018,Turner2018Oct,Serbyn2021Jun,Moudgalya2022,Chandran2023Mar}. Among these, we focus on the QMBS in this paper. 

QMBS states are low-entanglement but high-energy eigenstates of nonintegrable Hamiltonians and typically exhibit equally space energy levels. They give rise to atypical dynamics, such as long-lived or persistent revivals~\cite{Bernien2017,Turner2018,Turner2018Oct,Schecter2019Oct,Shibata2020May,Bluvstein2021controlling,Zhang2023,Su2023,Zhao2025}. On the theoretical side, a number of algebraic and symmetry-based frameworks have clarified the structural aspects of QMBS, including restricted spectrum generating algebra (RSGA)~\cite{batista2009canted,Mark2020,Moudgalya2020} and symmetry-based formalism ~\cite{ODea2020,Pakrouski2020Dec,Pakrouski2021Dec,Ren2021Mar,Ren2022Feb}.

Gotta {\it et al}. have proposed asymptotic quantum many-body scars (AQMBS)~\cite{Gotta2023}, which are closely related to QMBS but distinct from them. AQMBS states possess several characteristic properties: they are orthogonal to all QMBS states, have low entanglement, and, in contrast to QMBS, are not energy eigenstates of the Hamiltonian. Most importantly, their energy variance vanishes in the thermodynamic limit. As a result, AQMBS entail parametrically slow relaxation, characterized by diverging Mandelstam–Tamm bound~\cite{Mandelstam1945}. 

AQMBS states have been constructed in various models, including the spin-1 XY model~\cite{Schecter2019Oct,Gotta2023,PRXQuantum.5.040330}, the extended Fermi–Hubbard model with the correlated hopping terms~\cite{Mark2020,Moudgalya2020,kunimi2025}, the Heisenberg model with Dzyaloshinskii-Moriya interaction~\cite{Ren2024}, the domain-wall conserving model~\cite{Iadecola2020Jan,Ren2024}, the DH model~\cite{Kodama2023Jan,Kunimi2024Oct}, the PXP model~\cite{Turner2018,Turner2018Oct,Ren2024}, the Onsager scar model~\cite{Shibata2020May,Tamura2022,Ren2024}, the Affleck-Kennedy-Lieb-Tasaki model~\cite{Ren2024,wei2025spectroscopic}, and others~\cite{gotta2025open,gioia2025distinct}.

Building on the algebraic understanding of QMBS, a systematic route to AQMBS has recently been formulated~\cite{kunimi2025}: one identifies a parent Hamiltonian whose ground states coincide with the QMBS states, which are defined within a Hilbert subspace $\mathcal{H}_P$ containing the scar subspace $\mathcal{H}_{\rm scar}$. The AQMBS states can be obtained from the low-lying gapless excited states of the parent Hamiltonian provided that the system satisfies certain technical assumptions. This approach has been successfully applied to several benchmark models. In all examples, the resulting parent Hamiltonian is related to the spin-1/2 ferromagnetic Heisenberg model~\footnote{For some models, such as the DH model, the domain-wall conserving model, and the Onsager scar model, the parent Hamiltonian is not manifestly of the ferromagnetic Heisenberg type. In these cases, the parent Hamiltonians ultimately reduces to the ferromagnetic Heisenberg model after nontrivial mappings~\cite{Gomez1993,Cheong2009}.}. The AQMBS states are then identified as magnon-excited states associated with the Nambu-Goldstone mode.

The ubiquity of the spin-1/2 ferromagnetic Heisenberg parent naturally raises the question of whether different parent Hamiltonians can emerge within the same systematic construction. In this paper, we answer this question affirmatively by applying the AQMBS framework to $\mathrm{SU}(N)$ Hubbard models~\cite{Affleck1988Mar,Marston1989Jun,Honerkamp2004Apr,Hermele2009Sep,Cazalilla2009Oct,Gorshkov2010Apr,taie20126,scazza2014observation,Capponi2016Apr}, which are known to host QMBS via generalized $\eta$-pairing-type structures~\cite{Nakagawa2024}. We show that, upon projecting to the appropriate enlarged subspace and following the same construction steps, the parent Hamiltonian becomes the $\mathrm{SU}(N)$ ferromagnetic Heisenberg model. This yields a new family of parent Hamiltonians beyond the spin-$1/2$ case and demonstrates that higher-symmetry settings naturally emerge through the AQMBS construction, yielding gapless excitations in the form of $\mathrm{SU}(N)$ ferromagnetic magnons.

The rest of this paper is organized as follows: In Sec.~\ref{sec:framework}, we review the symmetry- and algebra-based framework underlying QMBS and its extension to AQMBS, fix notation, and present the multiladder generalization relevant to our setting. In Sec.~\ref{sec:results}, we apply this construction to the $\mathrm{SU}(N)$ Hubbard model: after introducing the model and the ladder operators, we derive the parent Hamiltonian and show that it is the $\mathrm{SU}(N)$ ferromagnetic Heisenberg chain. We then analyze the resulting AQMBS excitations, including their orthogonality to scar states, vanishing energy variance, dispersion, and entanglement scaling. Section~\ref{sec:summary} summarizes our findings and outlines open directions.

\section{General Framework}\label{sec:framework}
In this section, we summarize the algebraic and symmetry-based frameworks underlying QMBS and their extension to AQMBS. Throughout this paper, we assume lattice systems. We first formulate a multiladder extension of the restricted spectrum-generating algebra, then review the symmetry-based construction, and finally describe the systematic parent-Hamiltonian approach that will be applied to the SU($N$) Hubbard model in the next section.
\subsection{Extension of the restricted spectrum-generating algebra to the multiladder operator case}\label{subsec:RSGA}
To construct AQMBS states, the RSGA plays a central role~\cite{kunimi2025}. In previous works~\cite{batista2009canted,Moudgalya2020,Mark2020}, the RSGA is formulated for the case of a single ladder operator $\hat{Q}^{\dagger}$. In the SU($N$) Hubbard model, however, there exist $N-1$ independent ladder operators, which necessitates a generalization of the RSGA. In this subsection, we introduce the additional assumptions and notation required for the multiladder case, and summarize the resulting properties that will be used in later sections.

\subsubsection{Review of single ladder RSGA}

We briefly review the RSGA for single-ladder systems. Detailed proofs of Lemmas 1, 2, and 3 are given in Refs.~\cite{Moudgalya2020,Mark2020}.

Let $\hat{H}$ be a Hamiltonian and let $\ket{S_0}$ be a normalized eigenstate of $\hat H$ with eigenvalue $E_0$. We refer to $\ket{S_0}$ as a \textit{root eigenstate}, meaning that it serves as the starting point of the ladder construction introduced below. 
We consider a non-Hermitian operator $\hat Q^\dagger$ acting as a raising operator on $\ket{S_0}$. By definition of the root eigenstate, $\ket{S_0}$ cannot be obtained by acting with $\hat Q^\dagger$ on any other eigenstate of $\hat H$. We define a sequence of unnormalized states
\begin{align}
    \ket{\tilde{S}_n}\coloneq (\hat{Q}^\dagger)^n\ket{S_0}, \quad n=0,1,2,\ldots.\label{eq:definition_of_Sn_appendix}
\end{align}
\begin{lemma}[Spectrum generating algebra (SGA)]
    \label{lemma1}
    Let the Hamiltonian $\hat H$ satisfy the following conditions:
    \begin{align}
        \hat H\ket{S_0}&=E_0\ket{S_0},\label{eq:lemma1_condition1}\\
        [\hat H,\hat Q^\dagger]&=\mathcal{E}\hat Q^\dagger.\label{eq:lemma1_condition2}
    \end{align}
    Then, the following relation holds for all nonnegative integers $n\geq 0$:
    \begin{align}
        \label{eq:lemma1_statement}
        \hat H\ket{\tilde S_n}=(E_0+n\mathcal{E})\ket{\tilde S_n}, \quad \text{or}\quad \ket{\tilde S_n}=0.
    \end{align}
\end{lemma}
\begin{lemma}[RSGA-1]
    \label{lemma2}
    Let the Hamiltonian $\hat{H}$, the operator $\hat{Q}^{\dagger}$, and the state $\ket{S_0}$ satisfy the following conditions, with
    $\hat{Q}^{\dagger}\ket{S_0}\neq 0$:
    \begin{align}
        \hat{H}\ket{S_0}&=E_0\ket{S_0},\label{eq:RSGA1_condition1}\\
        [\hat{H},\hat{Q}^{\dagger}]\ket{S_0}&=\mathcal{E}\hat{Q}^{\dagger}\ket{S_0},\label{eq:RSGA1_condition2}\\
        [[\hat{H},\hat{Q}^{\dagger}],\hat{Q}^{\dagger}]&=0.\label{eq:RSGA1_condition3}
    \end{align}
    The following relation holds for all nonnegative integers $n\geq 0:$
    \begin{align}
        \hat{H}\ket{\tilde{S}_n}=(E_0+n\mathcal{E})\ket{\tilde{S}_n}\quad \text{or}\quad \ket{\tilde{S}_n}=0.\label{eq:RSGA1_statement}
    \end{align}
\end{lemma}
\begin{lemma}[RSGA-$m$]
    \label{lemma3}
    Let the Hamiltonian $\hat{H}$, the operator $\hat{Q}^{\dagger}$, and the state $\ket{S_0}$ satisfy the following conditions, assuming that $(\hat{Q}^{\dagger})^n\ket{S_0}\neq 0$ for $n\leq m$:
    \begin{align}
        \hat{H}\ket{S_0}&=E_0\ket{S_0},\label{eq:RSGA_M_condition1}\\
        [\hat{H},\hat{Q}^{\dagger}]\ket{S_0}&=\mathcal{E}\hat{Q}^{\dagger}\ket{S_0},\label{eq:RSGA_M_condition2}\\
        \underbrace{[[[\hat{H},\hat{Q}^{\dagger}],\hat{Q}^{\dagger}],\ldots]}_{r\text{ times}}\ket{S_0}&=0,\quad \text{for}\quad 2\leq r\leq m,\label{eq:RSGA_M_condition3}\\
        \underbrace{[[[\hat{H},\hat{Q}^{\dagger}],\hat{Q}^{\dagger}],\ldots]}_{m+1\text{ times}}&=0.\label{eq:RSGA_M_condition4}
    \end{align}
    The following relation holds for all nonnegative integers $n\geq 0$:
    \begin{align}
        \hat{H}\ket{\tilde S_n}=(E_0+n\mathcal{E})\ket{\tilde{S}_n}\quad \text{or}\quad \ket{\tilde S_n}=0.\label{eq:RSGA_M_statement}
    \end{align}
\end{lemma}

\subsubsection{Multiladder extension of RSGA}
We extend the RSGA from the single-ladder case to systems with multiple ladder operators, which we refer to as the multiladder RSGA-$m$ (MLRSGA-$m$). We consider a nonintegrable Hamiltonian $\hat{H}$ and a finite set of non-Hermitian operators $\{\hat{Q}_l^{\dagger}\}_{l=1}^{N}$. 
We assume the existence of a normalized reference state $\ket{S_{\bm 0}}$. The operators $\hat Q^\dagger_l$ act as independent raising operators on $\ket{S_{\bm 0}}$, and we assume that $\hat{Q}_l^{\dagger}\ket{S_{\bm 0}}\neq 0$ for all $l$. In analogy with the single-ladder case, we further assume that the reference state $\ket{S_{\bm 0}}$ cannot be obtained by acting with $\hat{Q}^{\dagger}_l$ on any eigenstate of $\hat H$, so that it serves as the root of the ladder construction.

Under these assumptions, the Hamiltonian $\hat{H}$ is said to satisfy the MLRSGA-$m$ conditions if it satisfies the following relations: 
\begin{align}
    \hat{H}\ket{S_{\bm 0}}&=E_0\ket{S_{\bm 0}},\label{eq:RSGA1}\\
    [\hat{H},\hat{Q}_l^{\dagger}]\ket{S_{\bm 0}}&=\mathcal{E}_l \hat Q_l^\dagger\ket{S_{\bm 0}},\label{eq:RSGA2}\\
    \underbrace{[[[[\hat{H},\hat{Q}_l^{\dagger}],\hat{Q}_l^{\dagger}],\ldots],\ldots]}_{r\text{ times}}\ket{S_{\bm 0}}&=0,\quad \text{for } 2\leq r\leq m,\label{eq:RSGA3}\\
    \underbrace{[[[[\hat{H},\hat{Q}_l^{\dagger}],\hat{Q}_l^{\dagger}],\ldots],\ldots]}_{m+1\text{ times}}&=0,\label{eq:RSGA4}\\
    [\hat{Q}^{\dagger}_l,\hat{Q}^{\dagger}_{k}]&=0,\quad \forall l, k,\label{eq:RSGA5}\\
    [[\hat{H},\hat{Q}^{\dagger}_l],\hat{Q}^{\dagger}_k]&=0\quad \forall l\neq k.\label{eq:RSGA6}
\end{align}
Here $E_0$ and $\mathcal{E}_l$ are real constants. 

Under the MLRSGA-$m$ conditions, the action of each ladder operator $\{\hat Q_l^\dagger\}_{l=1}^{N}$ on the scar states generated from $\ket{S_{\bm 0}}$ includes an energy shift $\mathcal{E}_l$. 
More precisely, for the multi-index $\bm m=(m_1,\ldots,m_N)$, we define the unnormalized scar states:
\begin{align}
    \ket{\tilde S_{\bm m}}&\coloneq
        \vspace{0.4em}\displaystyle{\prod_{l=1}^{N}(\hat Q_l^\dagger)^{m_l}\ket{S_{\bm 0}}},
\end{align}
where we consider the case $\ket{\tilde{S}_{\bm m}}\neq 0$.
Using Eqs.~\eqref{eq:RSGA1}-\eqref{eq:RSGA6}, one obtains
\begin{align}
    \label{eq:energy_tower}
    \hat H\ket{\tilde S_{\bm m}}&=\left(E_0+\sum_l m_l\mathcal{E}_l\right)\ket{\tilde S_{\bm m}}.
\end{align}
\begin{proof}[Proof of Eq.~\eqref{eq:energy_tower}]
    First, we focus on a single fixed mode $l$. In this case, Eqs.~\eqref{eq:RSGA1}-\eqref{eq:RSGA4} coincide with the assumptions of Lemma 3 upon replacing $\hat{Q}^{\dagger}$ with $\hat{Q}^{\dagger}_l$. Therefore, for any $m_l$ the following relation holds:
    \begin{align}
        \hat H(\hat Q_l^\dagger)^{m_l}\ket{S_{\bm 0}}&=(E_0+m_l\mathcal{E}_l)(\hat Q_l^\dagger)^{m_l}\ket{S_{\bm 0}},\notag\\
        &\text{ or}\quad (\hat Q^\dagger_l)^{m_l}\ket{S_{\bm 0}}=0.\label{eq:single_mode_energy}
    \end{align}
    Extracting the action of the commutator $[\hat H,(\hat Q^\dagger_l)^{m_l}]$ from Eq. \eqref{eq:single_mode_energy}, we obtain
    \begin{align}
        [\hat H,(\hat Q^\dagger_l)^{m_l}]\ket{S_{\bm 0}}&=m_l\mathcal{E}_l(\hat Q^\dagger_l)^{m_l}\ket{S_{\bm 0}}.\label{eq:commutator_single_mode}
    \end{align}

Next, we consider the action on the multimode state $\ket{\tilde{S}_{\bm m}}$. The commutator between $\hat H$ and a product of operators can be evaluated by repeated application of the Leibniz rule, $[\hat X,\hat Y\hat Z]=[\hat X,\hat Y]\hat Z+\hat Y[\hat X,\hat Z].$
    Iterating this identity yields,
    \begin{align}
        [\hat H,\hat A_1\hat A_2\cdots \hat A_N]&=\sum_{l=1}^N \hat A_1\cdots \hat A_{l-1}[\hat H,\hat A_l]\hat A_{l+1}\cdots \hat A_N,
    \end{align}
    where $\hat{A}_l$ is an arbitrary operator. The additional condition \eqref{eq:RSGA6} implies that $[\hat{H},\hat{Q}_l^{\dagger}]$ commutes with $\hat{Q}_k^{\dagger}$ for $k\neq l$, and the mutual commutativity of the ladder operators, $[\hat Q_l^\dagger,\hat Q_k^\dagger]=0$, is guaranteed by condition~\eqref{eq:RSGA5}. As a consequence, the commutator $[\hat H,(\hat Q_l^\dagger)^{m_l}]$ commutes with $(\hat Q_k^\dagger)^{m_k}$ for all $k\neq l$, allowing us to freely reorder the operators. Therefore, we obtain
    \begin{align}
        [\hat H,\prod_{n=1}^N(\hat Q^\dagger_n)^{m_n}]&=\sum_{l=1}^N \left(\prod_{n\neq l}(\hat Q^\dagger_n)^{m_n}\right)[\hat H,(\hat Q^\dagger_l)^{m_l}].
    \end{align}
    Using this identity, we prove Eq.~~\eqref{eq:energy_tower}:
    \begin{align}
        \hat{H}\ket{\tilde{S}_{\bm m}}&=\left[\hat{H},\prod_{l=1}^N(\hat{Q}^{\dagger}_l)^{m_l}\right]\ket{S_{\bm 0}} + \prod_{l=1}^N(\hat{Q}^{\dagger}_l)^{m_l}\hat{H}\ket{S_{\bm 0}}\notag\\
        &=\sum_{l=1}^N m_l\mathcal{E}_l\prod_{k=1}^N(\hat{Q}^{\dagger}_k)^{m_k}\ket{S_{\bm 0}} + E_0\prod_{l=1}^N(\hat{Q}^{\dagger}_l)^{m_l}\ket{S_{\bm 0}}\notag\\
        &=\left(E_0+\sum_{l=1}^N m_l\mathcal{E}_l\right)\ket{\tilde S_{\bm m}}.
    \end{align}
\end{proof}
Finally, we introduce the normalized scar state
\begin{align}
    \ket{S_{\bm m}}&\coloneq \frac{\ket{\tilde S_{\bm m}}}{\sqrt{\langle\tilde{S}_{\bm{m}}|\tilde{S}_{\bm{m}}\rangle}}.
\end{align}
Thus the states $\ket{S_{\bm m}}$ constitute a set of QMBS states.

\subsection{Symmetry-based formalism}\label{subsec:symmetry_based}
We now specialize the symmetry-based formalism to lattice systems, where local degrees of freedom are labeled by a site index $j$. O'Dea {\it et al}.~\cite{ODea2020} proposed a symmetry-based formalism that elucidates the structure of Hamiltonians supporting QMBS states. Their key observation is that such Hamiltonians can be systematically constructed from symmetry-preserving building blocks, providing a unifying reinterpretation of previously known models. In this subsection, we summarize the central aspects of this formalism, which will be used in the systematic construction presented in the next subsection.

We begin by fixing a semisimple Lie algebra $G$~\footnote{In the case of $\hat{C}=0$, Lie algebra $G$ need not be semisimple. } and, on each site $j$, introducing a Cartan subalgebra (CSA) generated by a maximal set of mutually commuting Hermitian operators $\{\hat{Q}^{z}_{\mu,j}\}\,(\mu=1,\dots,R)$ with $R=\mathrm{rank}(G)$. The remaining on-site generators are reorganized into root-labeled ladder operators ${\hat Q_{\bm{\alpha},j}}$ such that
\begin{align}
    [\hat{Q}^z_{\mu,j},\hat{Q}_{\bm{\alpha},j}]=\alpha_\mu\hat{Q}_{\bm{\alpha},j},
\end{align}
where $\bm\alpha=(\alpha_1,\ldots,\alpha_R)$ is a root vector of $G$. We denote positive and negative roots by $\bm \alpha^{\pm}$ such that, $\{\bm \alpha\}=\{\bm\alpha^+\}\cup\{\bm\alpha^-\}$ with $\bm{\alpha}^+\equiv-\bm{\alpha}^-$, so that 
$\hat Q_{\bm \alpha}$ correspond to ladder operators. We define the corresponding global operators as
\begin{align}
    \hat{Q}_{\mu}^{z}&\coloneq\sum_j \hat{Q}_{\mu,j}^{z},\\
    \hat{Q}_{\bm \alpha}&\coloneq\sum_j\hat{Q}_{\bm \alpha,j},
\end{align}
where $\sum_j$ runs over all lattice sites.

Within this symmetry-based formalism, we decompose the Hamiltonian as
\begin{align}
\hat{H}=\hat{H}_{\rm A}+\hat{H}_{\rm SG}+\hat{H}_{\rm sym}.\label{eq:Hamiltonian_symmetry_based_formalism}
\end{align}
The first term $\hat H_{\rm A}$ annihilates all scar states:
\begin{align}
\hat H_{\rm A}\ket{S_{\bm m}}=0, \quad \forall \bm m.\label{eq:definition_of_H_A}
\end{align}
The spectrum-generating term $\hat H_{\rm SG}$ is taken as
\begin{align}
\hat H_{\rm SG}=\sum_{\mu=1}^R h_\mu \hat Q_\mu^z + \hat C.\label{eq:definition_of_H_SG}
\end{align}
Here, $h_\mu\,(\mu=1,\ldots,R)$ are real coefficients with dimensions of energy that specify the spectrum generated by the root ladders through
\begin{align}
\mathcal{E}_{\bm{\alpha}}\coloneq \sum_{\mu=1}^R h_\mu \alpha_\mu,\label{eq:definition_of_mathcal_E}
\end{align}
for a root vector $\bm \alpha$. 
Since positive and negative roots satisfy $\bm \alpha^+=-\bm \alpha^-$, it follows that $\mathcal{E}_{\bm \alpha^+}=-\mathcal{E}_{\bm \alpha^-}$.
The hermitian operator $\hat C$ commutes with the CSA but generally does not commute with $\hat{Q}_{\bm{\alpha}}$:
\begin{align}
[\hat C,\hat{Q}_\mu^z]&=0, \quad \forall \mu,\label{eq:commutation_relation_C_and_Qz}\\
[\hat{C}, \hat{Q}_{\bm{\alpha}}]&\not=0.\label{eq:commutation_relation_C_and_Q_alpha_pm}
\end{align}
With this choice, the CSA part of $\hat H_{\rm SG}$ obeys the SGA,
\begin{align}
\left[~\sum_{\mu=1}^R h_\mu \hat{Q}_\mu^z,\hat{Q}_{\bm{\alpha}}\right]&=\mathcal{E}_{\bm \alpha} \hat{Q}_{\bm \alpha},
\end{align}
whereas for the full $\hat H_{\rm SG}$ one has
\begin{align}
[\hat H_{\rm SG},\hat Q_{\bm \alpha}]&=\mathcal{E}_{\bm \alpha} \hat Q_{\bm \alpha} + [\hat C,\hat Q_{\bm \alpha}].\label{eq:commutation_relation_HSG_and_Q_alpha_pm}
\end{align}

To account for the the presence of a noncommuting term $\hat C$, we show that $[\hat{C}, \hat{Q}_{\bm{\alpha}}]$ is itself a ladder operator with the same root $\bm{\alpha}$.
For $X=\hat Q_\mu^z, Y=\hat C$, and $Z=\hat Q_{\bm \alpha}$, the Jacobi identity $[X,[Y,Z]]+[Y,[Z,X]]+[Z,[X,Y]]=0$ gives
\begin{align}
[\hat{Q}_{\mu}^z, [\hat{C}, \hat{Q}_{\bm{\alpha}}]]=[\hat C,[\hat Q^z_\mu,\hat Q_{\bm \alpha}]]-[\hat Q_{\bm \alpha},[\hat Q^z_\mu,\hat C]].
\label{eq:Jacobi_identity}
\end{align}
Using the assumptions $[\hat{Q}_{\mu}^{z}, \hat{C}]=0$ and the definition of the ladder operator $[\hat{Q}_{\mu}^{z}, \hat{Q}_{\bm \alpha}]=\alpha_{\mu} \hat{Q}_{\bm{\alpha}}$, we obtain
\begin{align}
[\hat{Q}_{\mu}^z, [\hat{C}, \hat{Q}_{\bm{\alpha}}]] =\alpha_{\mu} [\hat{C}, \hat{Q}_{\bm{\alpha}}].\label{eq:commutator_from_Jacobi_idenfity}
\end{align}
For a simple Lie algebra, the root space for any nonzero root is one-dimensional, implying that any two ladder operators with the same root must be proportional. Therefore, one has $[\hat C, \hat Q_{\bm \alpha}] \propto \hat Q_{\bm \alpha}$. Therefore, within the subspace $\mathcal{S}_{\bm \alpha}$ spanned by products of root ladders acting on a seed state $\ket{\Omega}$, chosen to be the highest-weight state for a given root, we must have $[\hat C, \hat Q_{\bm \alpha}] \propto \hat Q_{\bm \alpha}$. This ensures that the action of the commutator remains within $\mathcal{S}_{\bm \alpha}$.
Consequently, we can work with a projected spectrum-generating algebra:
\begin{align}
\hat{\mathcal{P}}_{\mathcal{S}_{\bm\alpha}}[\hat H_{\rm SG},\hat Q_{\bm\alpha}]\hat{\mathcal{P}}_{\mathcal{S}_{\bm \alpha}}
&=(\mathcal{E}_{\bm \alpha}+\lambda_{\bm \alpha}) \hat{\mathcal{P}}_{\mathcal{S}_{\bm \alpha}}\hat Q_{\bm \alpha}\hat{\mathcal{P}}_{\mathcal{S}_{\bm \alpha}},
\end{align}
where $\hat{\mathcal{P}}_{\mathcal{S}_{\bm \alpha}}$ denotes the projection operator onto $\mathcal S_{\bm \alpha}$, and $\lambda_{\boldsymbol\alpha}$ is determined by the action of $\hat C$ within $\mathcal S_{\bm \alpha}$. Therefore, the presence of $\hat C$ does not break the ladder structure associated with each root.

The third term $\hat H_{\mathrm{sym}}$ of Eq.~(\ref{eq:Hamiltonian_symmetry_based_formalism}) commutes with all CSA and ladder generators (it can be set to zero) and therefore does not affect the spectrum-generating structure.

\subsection{Systematic construction of AQMBS: review and multiladder extension}\label{subsec:aqmbs}
In this subsection, we review the systematic construction of AQMBS proposed by Kunimi {\it et al.}~\cite{kunimi2025}, and extend the formalism to the case where the RSGA admits multiple ladder operators. The basic idea is to construct a parent Hamiltonian $\hat H_{\rm p}$ whose ground states coincide with the QMBS states of the original Hamiltonian; if $\hat H_{\rm p}$ is gapless, the low-energy gapless excited states can be regarded as AQMBS states of the original Hamiltonian.

First, we consider the total Hilbert space $\mathcal{H}$ and the Hamiltonian of the system $\hat{H}$. We assume that the system has QMBS states that satisfy the MLRSGA conditions, and that the Hamiltonian $\hat{H}$ can be written in the symmetry-based formalism. Following Ref.~\cite{kunimi2025}, we introduce the subspace $\mathcal{H}_P$ as the direct sum of mutually disjoint subspaces labeled by the multiindex $\bm m = (m_1,\ldots,m_N)$,
\begin{align}
\mathcal{H}_P&=\bigoplus_{\bm m\in \mathcal{I}}\mathcal{H}_{P_{\bm m}},\label{eq:definition_of_H_P}
\end{align}
with $m_i=0,1,\ldots$. The index set $\mathcal{I}$ is defined as
\begin{align}
\mathcal{I}&\coloneq\{ {\bm m} \mid \ket{S_{\bm m}}\neq 0\},\label{eq:definition_of_set_I}
\end{align}
and each Hilbert subspace is defined by
\begin{align}
\mathcal{H}_{P_{\bm m}}&\coloneq\mathrm{Span}\{\ket{\bm l}\in \mathcal{H} \mid \braket{\bm l}{S_{\bm m}}\neq 0\}.\label{eq:definition_of_H_P_m}
\end{align}
Here, $\{\ket{\bm l}\}$ denotes a set of product states taken from a simultaneous eigenbasis of the Cartan generators $\{\hat{Q}_{\mu}^z\}_{\mu=1}^R$. Owing to the assumed properties of $\hat{Q}^z_{\mu}$ and $\hat{Q}_n^{\dagger}$, states in different subspaces $\mathcal{H}_{P_{\bm m}}$ are orthogonal.

We now impose the following structural assumption on the annihilating part of the Hamiltonian $\hat{H}_{\rm A}$. We decompose it as
\begin{align}
\hat H_{\rm A}&\coloneq \hat{H}_0 + \hat{H}_{\rm p}',\label{eq:definition_of_H_A}\\
\hat H_0&\coloneq \sum_j \hat{h}_j,\label{eq:definition_of_H0}\\
\hat H_{\rm p}'&\coloneq \sum_{\bm m \in \mathcal{I}} c_{\bm m} \hat{\mathcal{P}}_{\bm m}\hat{H}_0^2\hat{\mathcal{P}}_{\bm m},\label{eq:definition_of_H_p'}
\end{align}
where $\hat{\mathcal{P}}_{\bm m}$ denotes the projection operator onto the subspace $\mathcal{H}_{P_{\bm m}}$ defined above, and $c_{\bm m}$ is a real constant with dimensions of the inverse of the energy. We assume that $\hat{h}_j$ is a local operator, which has support only in the vicinity of site $j$. Let $\hat{\mathcal{P}}\coloneq \sum_{\bm{m}\in \mathcal{I}}\hat{\mathcal{P}}_{\bm m}$ denote the projector onto the subspace $\mathcal{H}_P$. Additionally, we introduce the complement of $\mathcal{H}_P$ as $\mathcal{H}_Q$, and define the projection operator onto $\mathcal{H}_Q$ as $\hat{\mathcal{Q}}\coloneq \hat{1}-\hat{\mathcal{P}}$, where $\hat{1}$ is the identity operator in $\mathcal{H}$. Following Ref.~\cite{kunimi2025}, we impose the key structural assumption that each local term annihilates $\mathcal{H}_P$ after projection:
\begin{align}
\hat{\mathcal{P}}\hat h_j\hat{\mathcal{P}}&=0,\quad \forall j.
\end{align}
Physically, any single local action $\hat h_j$ takes states out of $\mathcal{H}_P$.
This assumption underlies the parent-Hamiltonian construction below.

We define $\hat{H}_{\rm p}$ as
\begin{align}
\hat{H}_{\rm p}&\coloneq \hat{\mathcal{P}}\hat{H}_0^2\hat{\mathcal{P}}.\label{eq:definition_of_parent_Hamiltonian}
\end{align}
We find that $\hat{H}_{\rm p}$ is positive semidefinite on $\mathcal{H}_P$ and annihilates all QMBS $\{\ket{S_{\bm m}}\}$ of the original Hamiltonian:
\begin{align}
\hat{H}_{\rm p}\ket{S_{\bm m}}&=0, \quad \forall \bm{m}\in \mathcal{I}.\label{eq:H_P_annihilate_scar}
\end{align}
Therefore, the ground-state manifold of $\hat H_{\rm p}$ is $\mathrm{Span}\{\ket{S_{\bm m}}\mid\bm m\in \mathcal{I}\}$, that is, $\hat H_{\rm p}$
is a parent Hamiltonian whose ground states are the QMBS of the original Hamiltonian.

From the definition of the parent Hamiltonian above, we can write
\begin{align}
    \hat H_{\rm p}&=\sum_{\bm n,\bm m \in \mathcal{I}}\hat{\mathcal{P}}_{\bm n}\hat H_0^2\hat{\mathcal{P}}_{\bm m}.
\end{align}
We now assume that the off-diagonal elements of the parent Hamiltonian vanish,
\begin{align}
\hat{\mathcal{P}}_{\bm n}\hat{H}_0^2\hat{\mathcal{P}}_{\bm m}&=0, \quad \forall \bm n\neq \bm m,\label{eq:assumption_of_off-diagonal_elements_Parent_Hamiltonian}
\end{align}
so that the parent Hamiltonian reduces to a sum over the Hilbert subspaces $\mathcal{H}_{P_{\bm m}}$:
\begin{align}
\hat{H}_{\rm p}&=\sum_{\bm{m} \in \mathcal{I}}\hat{\mathcal{P}}_{\bm m}\hat{H}_0^2\hat{\mathcal{P}}_{\bm m}\eqcolon\sum_{\bm{m}\in\mathcal{I}}\hat{H}_{\rm p}^{(\bm{m})}.\label{eq:parent_Hamiltonian_only_diagonal_term}
\end{align}

We quantify the quality of candidate AQMBS states by the energy variance within each subspace $\mathcal H_{P_{\bm m}}$. For a normalized state 
$|\phi_{\bm m}\rangle\in\mathcal H_{P_{\bm m}}$, we define
\begin{align}
\Delta E_{\bm m}^2&\coloneq \bra{\phi_{\bm m}}\hat H^2\ket{\phi_{\bm m}} - (\bra{\phi_{\bm m}}\hat H\ket{\phi_{\bm m}})^2\notag \\
&=\bra{\phi_{\bm m}}\hat{\mathcal{P}}_{\bm m}\hat H^2\hat{\mathcal{P}}_{\bm m}\ket{\phi_{\bm m}} - (\bra{\phi_{\bm m}}\hat{\mathcal{P}}_{\bm m}\hat H\hat{\mathcal{P}}_{\bm m}\ket{\phi_{\bm m}})^2,\label{eq:calculation_energy_variance}
\end{align}
where we used $\hat{\mathcal{P}}_{\bm m}\ket{\phi_{\bm m}}=\ket{\phi_{\bm m}}$. Employing the relations established in this subsection, such as $\hat{\mathcal{P}}_{\bm{m}}\hat{h}_j\hat{\mathcal{P}}_{\bm{m}}=0$, $\hat{\mathcal{P}}_{\bm{m}}(\hat{H}_{\rm p}'+\hat{H}_{\rm SG}+\hat{H}_{\rm sym})\hat{\mathcal{Q}}=0$, and $\hat{\mathcal{Q}}(\hat{H}_{\rm p}'+\hat{H}_{\rm SG}+\hat{H}_{\rm sym})\hat{\mathcal{P}}_{\bm{m}}=0$, we find that the energy variance reduces to a contribution involving only $\hat{H}_0$
when $|\phi_{\bm m}\rangle$ is chosen as an eigenstate of $\hat{\mathcal{P}}_{\bm m} \hat H_0^{2}\hat{\mathcal{P}}_{\bm m}$.
Concretely, the energy variance becomes
\begin{align}
\Delta E_{\bm m}^2&=\bra{AS_{\bm m}}\hat{\mathcal{P}}_{\bm m}\hat H_0^2\hat{\mathcal{P}}_{\bm m}\ket{AS_{\bm m}},\label{eq:energy_variance_final_form}
\end{align}
where $|AS_{\bm m}\rangle$ is a normalized eigenstate of $\hat{\mathcal{P}}_{\bm m} \hat H_0^{2}\hat{\mathcal{P}}_{\bm m}$. Consequently, if $\hat H_{\rm p}^{(\bm m)}$ is gapless in the thermodynamic limit, its low-energy gapless excited states have vanishing energy variance and constitute AQMBS of the original Hamiltonian.

As shown in Ref.~\cite{kunimi2025}, the parent Hamiltonian reduces to the spin-$1/2$ ferromagnetic Heisenberg model in many cases. In such cases, the low-energy gapless excited states (i.e., magnon-excited states) can be obtained analytically from simple algebraic relations. However, the systematic construction does not guarantee that the parent Hamiltonian always reduces to a pure ferromagnetic Heisenberg model. When this is not the case, we need to rely on approximations or numerical calculations to obtain the AQMBS.

\section{Results: Application to SU($N$) Hubbard Model}\label{sec:results}
Previous applications of the parent-Hamiltonian scheme have yielded the spin-$1/2$ ferromagnetic Heisenberg model as described in the previous section~\cite{kunimi2025}. In what follows, we apply the formalism developed in the previous section to the SU($N$) Hubbard model, which exhibits higher internal symmetry and supports multiple ladder operators. We show that the parent Hamiltonian in this setting is the ferromagnetic SU($N$) Heisenberg model, and the AQMBS states are identified with gapless excited states of this parent Hamiltonian.

\subsection{Model}\label{subsec:Model}
We consider the $L$-site one-dimensional SU($N$) Fermi Hubbard model with $N\geq 3$, where the $N$ fermionic flavors are labeled by $\alpha_1,\alpha_2,\ldots,\alpha_N$. The Hamiltonian is defined by
\begin{align}
\hat{H}&\coloneq\hat{H}_{\mathrm{hop}}+\hat{H}_{\mathrm{int}},\label{eq:definition_of_SU(N)_Hubbard_Hamiltonian}\\
\hat{H}_\mathrm{hop}&\coloneq-J\sum_j \sum_{n=1}^N\left(\hat{c}^{\dagger}_{j,\alpha_n}\hat{c}_{j+1,\alpha_n}+\hat{c}^{\dagger}_{j+1,\alpha_n} \hat{c}_{j,\alpha_n}\right),\label{eq:definition_of_hopping_term_SU(N)_Hubbard}\\
\hat{H}_\mathrm{int}&\coloneq\frac{U}{2}\sum_j \sum_{n,m=1,n\not=m}^N\hat n_{j,\alpha_m}\hat n_{j,\alpha_n},\label{eq:definition_of_onsite_interaction_SU(N)_Hubbard}
\end{align}
where $\hat{c}^{\dagger}_{j,\alpha_n}$ ($\hat{c}_{j,\alpha_n}$) creates (annihilates) a fermion of flavor $\alpha_n$ at site $j$, $\hat{n}_{j,\alpha_n}=\hat{c}^\dagger_{j,\alpha_n}\hat{c}_{j,\alpha_n}$ is the corresponding number operator, the hopping amplitude $J$ is taken to be real, and $U$ denotes the on-site interaction strength between fermions of different flavors. We assume either periodic or open boundary conditions, depending on the context.

It has been shown in Ref.~\cite{Nakagawa2024} that the SU($N$) Hubbard model supports an exact tower of QMBS states generated by $\eta$-pairing. We define the $\eta$-pairing creation operator for flavors $\alpha_m$ and $\alpha_n$ ($m\neq n$) as
\begin{align}
\hat{\eta}^\dagger_{m n}&\coloneq\sum_j\hat{\eta}^\dagger_{j,mn}\coloneq\sum_j (-1)^j\hat{c}^\dagger_{j,\alpha_m}\hat{c}^\dagger_{j,\alpha_n}.\label{eq:eta_operator}
\end{align}
For the two-component Fermi-Hubbard model, Eq.~\eqref{eq:eta_operator} reduces to the $\eta$ operator originally introduced by Yang \cite{Yang1989}. In the notation of Sec.~II, the $\eta$-pairing operators $\hat{\eta}^\dagger_{m n}$ defined in Eq.~\eqref{eq:eta_operator} play the role of the RSGA ladder operators ${\hat Q_n^\dagger}$: each flavor pair $(\alpha_m,\alpha_n)$ corresponds to a single ladder direction $n$, and the vacuum $\vac$ serves as the reference state $\ket{S_{\bm{0}}}$.

Starting from the vacuum state $\vac$, we can construct a family of QMBS states labeled by the multiindex $\bm{m}'\equiv (m_2',m_3',\ldots,m_N')$ as
\begin{align}
&\ket{\tilde{S}'_{\bm{m}'}}\coloneq\prod_{n=2}^N\left(\hat{\eta}_{1,n}^\dagger\right)^{m_n'}\vac,\label{eq:scar_state_of_SU(N)_Hubbard}
\end{align}
where the nonnegative integers $m_n'$ ($n=2,3,\ldots,N$) satisfy $\sum_{n=2}^N m_n'\leq L$ and specify the number of $\eta$ pairs of each flavor combination ($\alpha_1,\alpha_n$). These states and operators $\hat{\eta}^\dagger_{m n}$ satisfy the MLRSGA-1 condition. The state $\ket{\tilde{S}'_{\bm{m}'}}$ satisfies
\begin{align}
\hat{H}\ket{\tilde{S}'_{\bm{m}'}}=U\sum_{n=2}^Nm_n'\ket{\tilde{S}'_{\bm{m}'}},\label{eq:eigenvalue_eq_SU(N)_Hubbard}
\end{align}
and hence the states $\{\ket{\tilde{S}'_{\bm{m}'}}\}$ constitute an exact tower of QMBS states in this model. 

Although the QMBS states can be constructed using the procedure described above, it is convenient to introduce a unitary transformation $\hat{U}$ and work with the unitary-transformed Hamiltonian, which simplifies the subsequent analysis. The unitary operator $\hat{U}$ is defined as
\begin{align}
\hat{U}&\coloneq\prod_{j\in A}\hat{U}_j,\label{eq:definition_of_total_unitary_operator_chang_sign}\\
\hat{U}_j&\coloneq\exp\left(i\pi\hat{n}_{j,\alpha_1}\right),\label{eq:unitary}
\end{align}
where we introduce the bipartition of the system into sublattices $A=\{1,3,\dots\}$ and $B=\{2,4,\dots\}$. The unitary-transformed Hamiltonian is then given by
\begin{align}
\hat{H}'\coloneq \hat{U}\hat{H}\hat{U}^{\dagger}=\hat{U}\hat{H}_{\rm hop}\hat{U}^{\dagger}+\hat{H}_{\rm int},\label{eq:definition_of_unitary_transformed_Hamiltonian}
\end{align}
where we used the fact that the on-site interaction term is invariant under the unitary transformation. Using the properties $\hat{U}_{j}\hat{c}_{j,\alpha_1}\hat U_{j}^\dagger=-
\hat{c}_{j,\alpha_1}$ and $\hat{U}_{j}\hat{c}_{j,\alpha_n}\hat{U}_{j}^\dagger=\hat{c}_{j,\alpha_n}$ for $n\geq 2$, we obtain the transformed hopping Hamiltonian:
\begin{align}
\hat{H}_{\rm hop}'&\coloneq\hat{U}\hat{H}_{\rm hop}\hat{U}^{\dagger}\notag \\
&=J\sum_j\Bigg[\hat{c}^{\dagger}_{j,\alpha_1}\hat{c}_{j+1,\alpha_1}+\hat{c}^{\dagger}_{j+1,\alpha_1}\hat{c}_{j,\alpha_1} \notag \\
&\left.\quad-\sum_{n=2}^N\left(\hat{c}^{\dagger}_{j,\alpha_n}\hat{c}_{j+1,\alpha_n}+\hat{c}^{\dagger}_{j+1,\alpha_n}\hat{c}_{j,\alpha_n}\right)   \right]\notag \\
&\eqcolon \sum_j\hat{h}_{j,j+1},\label{eq:definition_of_transformed_hopping_term}
\end{align}
where the hopping amplitude for the reference flavor $\alpha_1$ is $+J$, while for the other flavors it is $-J$. In the following, we take $\hat{H}'=\hat{H}_{\rm hop}'+\hat{H}_{\rm int}$ as the total Hamiltonian. Under the same unitary transformation, the $\eta$-pairing operators $\hat{\eta}^{\dagger}$ become
\begin{align}
\hat{Q}^{\dagger}_{m n}\equiv \hat{U}\hat{\eta}^{\dagger}_{m n}\hat{U}^{\dagger}=\sum_j\hat{c}^{\dagger}_{j,\alpha_m}\hat{c}^{\dagger}_{j,\alpha_n},\label{eq:definition_of_Q_dagger_SU(N)}
\end{align}
so that the factor $(-1)^j$ in Eq.~(\ref{eq:eta_operator}) is removed. Accordingly, the QMBS states are transformed as
\begin{align}
\ket{\tilde{S}_{\bm{m}'}}\coloneq \hat{U}\ket{\tilde{S}'_{\bm{m}'}}=\prod_{n=2}^N(\hat{Q}^{\dagger}_{1,n})^{m_{n}'}\ket{{\rm vac}}.\label{eq:transformed_QMBS}
\end{align}

Here, we briefly remark on the case $N=3$. According to Ref.~\cite{Nakagawa2024}, this case is distinct from $N\ge 4$ due to the presence of the additional exact eigenstates. Details of this case are discussed in the Appendix.

We also briefly comment on the nonintegrability of the model in Eq. (\ref{eq:definition_of_SU(N)_Hubbard_Hamiltonian}). It is well known that the one-dimensional SU($2$) Hubbard model is integrable~\cite{essler2005one}. In contrast to the case $N=2$, the SU($N$) Hubbard model for $N\ge 3$ is generically considered to be nonintegrable, except in special symmetry sectors such as the single-particle sectors and SU($2$) subsectors~\cite{Nakagawa2024}. Recently, it has also been proven that the $N=2$ Hubbard model in higher dimensions does not possess nontrivial local conserved quantities, which strongly supports its nonintegrability~\cite{futami2025absence}.

\subsection{Parent Hamiltonian}\label{subsec:parent_Hamiltonian}
For definiteness, we work with the one-dimensional SU($N$) Fermi-Hubbard model ($N\ge 3$) on a chain of length $L$. On a single site $j$, we denote the local vacuum by $\vac_j$ and introduce the on-site two-particle states as
\begin{align}
\hat c_{j,\alpha_m}^\dagger \hat c_{j,\alpha_n}^\dagger\vac_j\coloneq \ket{\alpha_m\alpha_n}_j,\quad (m<n).\label{eq:definition_of_local_eta_pair_state}
\end{align}
For later convenience, we set
\begin{align}
\ket{0}_j&\coloneq \vac_j,\label{eq:definition_of_0_state}\\
\ket{n-1}_j&\coloneq \ket{\alpha_1\alpha_n}_j,\quad (n=2,3,\ldots,N),\label{eq:definition_of_mu-1_state}
\end{align}
so that the local basis $\ket{l}_j$ consists of $N$ states $\ket{0}_j,\ket{1}_j,\ldots,\ket{N-1}_j$ built from the vacuum (holon, i.e., an empty on-site state) and the doublons (i.e., doubly occupied on-site states) that involve the reference flavor $\alpha_1$.

The Hilbert subspace with nonzero overlap with the scar subspace is defined as
\begin{align}
\mathcal{H}_P=\mathrm{Span}\{\ket{\bm \ell}=\bigotimes_{j=1}^L \ket{\ell_j}_j \mid \ell_j=0,1,\ldots,N-1\}.\label{eq:definition_of_subspace}
\end{align}
Within $\mathcal{H}_P$, configurations may be viewed as a gas of hardcore doublons with $N-1$ flavors $(\ell=1,\ldots,N-1)$ moving on an $L$-site background in the presence of holons $(\ell=0)$. This subspace will serve as the stage on which we construct a parent Hamiltonian whose exact ground-state manifold reproduces the tower of scarred states discussed above.

Within the constrained subspace $\mathcal H_P$, it is convenient to work with state-changing operators acting on the local basis ${|0\rangle_j,|1\rangle_j,\ldots,|N-1\rangle_j}$.
We define
\begin{align}
\hat F^{\mu\nu}_j&\coloneq \ket{\mu}_j\bra{\nu}_j, \quad \mu,\nu=0,1,\ldots,N-1,\label{eq:definition_of_F_munu}
\end{align}
so that $\hat F^{\mu\nu}_j$ maps $\ket{\nu}_j$ to $\ket{\mu}_j$ and annihilates states orthogonal to $\ket{\nu}_j$, i.e.,
\begin{align}
\hat F^{\mu\nu}_j\ket{\lambda}_j&=\delta_{\nu\lambda}\ket{\mu}_j,\label{eq:F_acting_of_lambda}\\
(\hat F^{\mu\nu}_j)^\dagger&=\hat F^{\nu\mu}_j.\label{eq:hermitian_conjugate_of_F_munu}
\end{align}
For example, $\hat F^{10}_j=\ket{1}_j\bra{0}_j$ converts the local state from a holon to a doublon of flavor $\alpha_1\alpha_2$. Operators on different sites commute, $[\hat F^{\mu\nu}_j,\hat F^{\kappa\lambda}_k]=0$ for $j\neq k$, while on the same site they satisfy the matrix-unit algebra~\cite{Cazalilla2014Nov}
\begin{align}
[\hat{F}^{\mu\nu}_j, \hat{F}^{\kappa\lambda}_j]&=\delta_{\nu\kappa}\hat{F}^{\mu\lambda}_j-\delta_{\mu\lambda}\hat{F}^{\kappa\nu}_j.\label{eq:commutation_relation_SU(N)}
\end{align}
Thus, the operators $\hat{F}_j^{\mu\nu}$ constitute the standard SU($N$) spin operators acting on $\mathcal H_P$. We will construct the parent Hamiltonian below in terms of bilinears of these operators.

In the symmetry-based formalism summarized in Sec.~\ref{subsec:symmetry_based}, the Hamiltonian of the SU($N$) Hubbard model is decomposed as $\hat{H}'=\hat H_{\rm A}+\hat{H}_{\rm SG}+\hat{H}_{\rm sym}$, where $\hat{H}_{\rm A}=\hat{H}_{0}=\hat{H}'_{\rm hop}$, $\hat{H}_{\rm SG}=\hat{H}_{\rm int}$, and $\hat{H}_{\rm sym}=0$. The local operator $\hat{h}_j$ in Eq.~(\ref{eq:definition_of_H0}) corresponds to $\hat{h}_{j,j+1}$ in Eq.~(\ref{eq:definition_of_transformed_hopping_term}). The parent Hamiltonian of the system can be written as
\begin{align}
\hat{H}_{\rm p}=\sum_{j,j'}\hat{\mathcal{P}}\hat{h}_{j,j+1}\hat{h}_{j',j'+1}\hat{\mathcal{P}}.\label{eq:parent_Hamiltonian_calculation1}
\end{align}
As shown in Appendix A of Ref.~\cite{kunimi2025}, if the Hilbert subspace $\mathcal{H}_P$ has a tensor product structure, the only nonvanishing terms in Eq.~(\ref{eq:parent_Hamiltonian_calculation1}) are those for which the supports of the operators $\hat{h}_{j,j+1}$ and $\hat{h}_{j',j'+1}$ overlap. Because the present system satisfies this tensor product condition, the parent Hamiltonian reduces to
\begin{align}
\hat{H}_{\rm p}=\sum_j\hat{\mathcal{P}}(\hat{h}_{j,j+1}^2+\hat{h}_{j,j+1}\hat{h}_{j+1,j+2}+\hat{h}_{j+1,j+2}\hat{h}_{j,j+1})\hat{\mathcal{P}}.\label{eq:parent_Hamiltonian_calculation2}
\end{align}
A direct calculation shows that the second and third terms in Eq.~(\ref{eq:parent_Hamiltonian_calculation2}) vanish because $\hat{h}_{j,j+1}\hat{h}_{j+1,j+2}\hat{\mathcal{P}}$ and $\hat{h}_{j+1,j+2}\hat{h}_{j,j+1}\hat{\mathcal{P}}$ always yield singlon or triplon (i.e., states with single or triple on-site occupancy), which are not elements of $\mathcal{H}_P$. Therefore, the parent Hamiltonian becomes
\begin{align}
\hat{H}_{\rm p}=\sum_j\hat{\mathcal{P}}\hat{h}_{j,j+1}^2\hat{\mathcal{P}}.\label{eq:parent_Hamiltonian_local_form}
\end{align}
For later convenience, we now introduce the local Hilbert subspace and its projector:
\begin{align}
\mathcal{H}_{P_{j,j+1}}&=\mathrm{Span}\{\ket{\ell}_j\otimes\ket{\ell'}_{j+1} ~|~\ell,\ell'=0,1,\ldots,N-1\},\label{eq:definition_of_local_Hilbert_space}\\
\hat{\mathcal{P}}_j&\coloneq\sum_{\ell=0}^{N-1}\ket{\ell}_j\bra{\ell}_j=\sum_{\ell=0}^{N-1} \hat F^{\ell\ell}_j.\label{eq:definition_of_local_projection_operator}
\end{align}
The two-site projector is defined by
\begin{align}
\hat{\mathcal{P}}_{j,j+1}&\coloneq\hat{\mathcal{P}}_j\hat{\mathcal{P}}_{j+1}\notag\\
&=\sum_{\mu=0}^{N-1}\sum_{\nu=0}^{N-1}\hat F^{\mu\mu}_j\hat F^{\nu\nu}_{j+1}\notag\\
&=\sum_{\mu,\nu=0}^{N-1}\ket{\mu}_j\ket{\nu}_{j+1}\bra{\mu}_j\bra{\nu}_{j+1}.\label{eq:definition_of_two_site_projection_operator}
\end{align}
By construction, $\hat{\mathcal{P}}_{j,j+1}^\dagger=\hat{\mathcal{P}}_{j,j+1}$ and $\hat{\mathcal{P}}_{j,j+1}^2=\hat{\mathcal{P}}_{j,j+1}$ hold. The parent Hamiltonian can be written as
\begin{align}
\hat{H}_{\rm p}=\sum_j\hat{\mathcal{P}}_{j,j+1}\hat{h}_{j,j+1}^2\hat{\mathcal{P}}_{j,j+1}.\label{eq:parent_Hamiltonian_bond_rep}
\end{align}

To obtain the explicit expression for the parent Hamiltonian, we investigate what happens when $\hat{h}_{j,j+1}$ acts on the Hilbert subspace $\mathcal{H}_{P_{j,j+1}}$. There are four possible states in $\mathcal{H}_{P_{j,j+1}}$: holon-holon ($\ket{{\rm vac}}_j\ket{{\rm vac}}_{j+1}=\ket{0}_j\ket{0}_{j+1}$), holon-doublon ($\ket{{\rm vac}}_j\ket{\alpha_1\alpha_{\mu}}_{j+1}=\ket{0}_{j}\ket{\mu-1}_{j+1}$), doublon-holon ($\ket{\alpha_1\alpha_{\mu}}_j\ket{{\rm vac}}_{j+1}=\ket{\mu-1}_j\ket{0}_{j+1}$), and doublon-doublon ($\ket{\alpha_1\alpha_{\mu}}_j\ket{\alpha_1\alpha_{\nu}}_{j+1}=\ket{\mu-1}_j\ket{\nu-1}_{j+1}$) states ($\mu,\nu\ge 2$). In the holon-holon case, we obtain $\hat{h}_{j,j+1}\ket{{\rm vac}}_j\ket{{\rm vac}}_{j+1}=0$. We next consider the holon-doublon and doublon-holon cases. Acting once with $\hat h_{j,j+1}$ takes doublon-holon or holon-doublon configuration out of $\mathcal{H}_{P_{j,j+1}}$ into the intermediate single-occupancy (singlon-singlon) states $\ket{\alpha_1}_j\ket{\alpha_{\mu}}_{j+1}$ and $\ket{\alpha_{\mu}}_j\ket{\alpha_1}_{j+1}$ ($\mu\ge 2$). Writing $\ket{X,Y}\equiv\ket{X}_j\ket{Y}_{j+1}$, we obtain
\begin{align}
\hat{h}^2_{j,j+1}\ket{\alpha_1\alpha_{\mu},{\rm vac}}&=\hat{h}_{j,j+1}\left(-J\ket{\alpha_1,\alpha_{\mu}}-J\ket{\alpha_{\mu},\alpha_1}\right)\notag\\
&=+2J^2\left(\ket{\alpha_1\alpha_{\mu},{\rm vac}}-\ket{{\rm vac},\alpha_1\alpha_{\mu}}\right),\label{eq:calculation_h_squared_to_doublon-holon_state}\\
\hat{h}^2_{j,j+1}\ket{{\rm vac},\alpha_1\alpha_{\mu}}&=\hat{h}_{j,j+1}\left(J\ket{\alpha_{\mu},\alpha_1}+J\ket{\alpha_1,\alpha_{\mu}}\right)\notag\\
&=-2J^2\left(\ket{\alpha_1\alpha_{\mu},{\rm vac}}-\ket{{\rm vac},\alpha_1\alpha_{\mu}}\right).\label{eq:calculation_h_squared_to_holon_doublon_state}
\end{align}
In the doublon-doublon case, we obtain
\begin{align}
&\hat h^2_{j,j+1}\ket{\alpha_1\alpha_{\mu},\alpha_1\alpha_{\nu}}\notag \\
&=\hat h_{j,j+1}\left(J\ket{\alpha_1,\alpha_1\alpha_{\mu}\alpha_{\nu}}+J\ket{\alpha_1\alpha_{\mu}\alpha_{\nu},\alpha_1}\right)\notag\\
&=2J^2\left(\ket{\alpha_1\alpha_{\mu},\alpha_1\alpha_{\nu}}-\ket{\alpha_1\alpha_{\nu},\alpha_1\alpha_{\mu}}\right).\label{eq:calculation_h_squared_to_doublon-doublon_state}
\end{align}
These identities establish the sign rule used below: two-hop ``round-trip" processes, in which a doublon virtually leaves a site and returns with the same flavor, contribute with a positive sign, whereas processes that exchange a doublon between the two sites (or swap two doublons) contribute with a negative sign.

Projecting the second-order bond operator onto the constrained subspace on $(j,j+1)$ gives
\begin{align}
\hat{\mathcal{P}}_{j,j+1}\hat h^2_{j,j+1}\hat{\mathcal{P}}_{j,j+1}&=2J^2\sum_{\substack{\mu,\nu=0\\\mu\neq\nu}}^{N-1}
\left(\hat{F}^{\mu\mu}_j\hat{F}^{\nu\nu}_{j+1}-\hat{F}^{\mu\nu}_j\hat{F}^{\nu\mu}_{j+1}\right),\label{eq:calculation_P_h_squared_P}
\end{align}
where the overall coefficient follows from the two-hop processes discussed above. Using the identities
\begin{align}
\sum_{\mu=0}^{N-1}\hat{F}^{\mu\mu}_{j}&=\hat{I}_j,\label{eq:Identify_sum_F}\\
\sum_{\substack{\nu=0\\\nu\neq\mu}}^{N-1}\hat{F}^{\mu\mu}_j\hat{F}^{\nu\nu}_{j+1}&=\hat{F}^{\mu\mu}_j(\hat{I}_{j+1}-\hat{F}^{\mu\mu}_{j+1}),\label{eq:Identify_product_F}
\end{align}
we can rewrite Eq.~(\ref{eq:calculation_P_h_squared_P}) as
\begin{align}
\hat{\mathcal{P}}_{j,j+1}\hat{h}^2_{j,j+1}\hat{\mathcal{P}}_{j,j+1}&=2J^2\left(\hat I_{j,j+1}-\sum_{\mu,\nu=0}^{N-1}\hat{F}^{\mu\nu}_j\hat{F}^{\nu\mu}_{j+1}\right).\label{eq:rewrite_P_h_squared_P}
\end{align}
Here, $\hat{I}_j$ is the identity operator acting on site $j$, and $\hat{I}_{j,j+1}\coloneq\hat{I}_j\hat{I}_{j+1}$. Summing over all nearest-neighbor bonds, we finally obtain the parent Hamiltonian
\begin{align}
\hat H_{\rm p}&=\sum_j\hat{\mathcal{P}}_{j,j+1}\hat{h}^2_{j,j+1}\hat{\mathcal{P}}_{j,j+1}\notag\\
&=2J^2\sum_j \left(\hat{I}_{j,j+1}-\sum_{\mu,\nu=0}^{N-1}\hat{F}^{\mu\nu}_j\hat{F}^{\nu\mu}_{j+1}\right).\label{eq:parent_Hamiltonian_final_form}
\end{align}
This is the ferromagnetic SU($N$) Heisenberg Hamiltonian up to a constant shift restricted to the subspace $\mathcal{H}_P$.

\subsection{Gapless excitations of the parent Hamiltonian}\label{subsec:gapless_excitations}
Having established that the parent Hamiltonian is equivalent to the SU($N$) ferromagnetic Heisenberg model, we now turn to its low-energy excitations. It is well known that the ferromagnetic Heisenberg model supports gapless magnon excitations~\cite{Auerbach1998interacting}, which naturally provide candidates for the AQMBS states in our framework. In the following, we first impose periodic boundary conditions and construct one-magnon excitations with well-defined momentum. Subsequently, we analyze the case of open boundary conditions, where the corresponding low-energy excitations are obtained by constructing standing-wave-like states. Similar calculations have been performed in Appendix C of Ref.~\cite{kunimi2025}.

We start from the parent Hamiltonian obtained in the previous subsection, which reduces to the SU($N$) ferromagnetic Heisenberg model on a ring of length $L$. Let us consider single–quasiparticle gapless excitations above the fully polarized reference state or the vacuum state in the fermionic language:
\begin{align}
\ket{\bm{0}}&\coloneq\bigotimes_{j=1}^L \ket{0}_j.\label{eq:definition_of_bm0_state}
\end{align}
To create a momentum eigenstate, we introduce the Fourier-transformed on-site ladder operator
\begin{align}
\hat{F}^{\mu0}_{\mathrm{PBC}}(k)&\coloneq\frac{1}{\sqrt{L}}\sum_{j=1}^L e^{ikj}\hat{F}^{\mu0}_j, \quad \mu=1,2,\ldots,N-1,\label{eq:definition_of_Fourier_transformed_ladder_operator}
\end{align}
where $k=2\pi n/L$ is a crystal momentum with an integer $n$.
By translational invariance, $\hat{F}^{\mu0}_{\mathrm{PBC}}(k)\ket{\bm{0}}$ carries a well-defined crystal momentum $k$. Using the commutation relations in the one-magnon sector, we obtain
\begin{align}
[\hat{H}_{\rm p},\hat{F}^{\mu0}_{\mathrm{PBC}}(k)]\ket{\bm{0}}&=\frac{1}{\sqrt{L}}\sum_{j=1}^L e^{ikj}[\hat{H}_{\rm p},\hat{F}^{\mu0}_j]\ket{\bm{0}}\notag\\
&=4J^2\left(1-\cos k\right)\hat{F}^{\mu0}_{\mathrm{PBC}}(k)\ket{\bm{0}}.\label{eq:RSGA_like_relation_for_SU(N)_Heisenberg_model}
\end{align}
Therefore, the one-quasiparticle state $\hat F^{\mu0}_{\mathrm{PBC}}(k)\ket{0}$ is an exact eigenstate in the one-magnon sector with dispersion
\begin{align}
\mathcal{E}(k)&\coloneq 4J^2(1-\cos k),\label{eq:definition_of_magnon_dispersion_PBC}
\end{align}
which is gapless and quadratic near $k=0$:
\begin{align}
\mathcal{E}(k)&\sim 2J^2k^2 \quad (k\to 0).\label{eq:dispersion_near_k=0_PBC}
\end{align}
Consequently, for $k=2\pi n/L$ with fixed $n=O(1)$, we have $\mathcal{E}(k)\to 0$ as $L\to\infty$.

We now turn to open boundary conditions. Taking the fully polarized state $\ket{\bm{0}}$ as the reference, we define the one-magnon basis for flavor $\mu~(=1,2,\ldots,N-1)$ by
\begin{align}
\left\{\ket{M_j^{\mu}}\coloneq\hat F^{\mu 0}_j\ket{\bm{0}} \mid j=1,2,\ldots,L\right\}.\label{eq:definition_of_one-magnon_basis_OBC}
\end{align}
For later use, we record the action of the parent Hamiltonian on a bond ($j,j+1$) as
\begin{align}
\hat{H}_{\rm p}^{(j,j+1)}=2J^2\sum_{\substack{\mu,\nu=0\\\mu\neq\nu}}^{N-1}\left(\hat{F}^{\mu\mu}_j\hat F^{\nu\nu}_{j+1}-\hat{F}^{\mu\nu}_j\hat{F}^{\nu\mu}_{j+1}\right).\label{eq:parent_Hamiltonian_bond_j_j+1_OBC}
\end{align}
Restricting $\hat H_{\rm p}^{(j,j+1)}$ to the one-magnon sector spanned by the two-site states $\{\ket{0,0},\ket{\mu,0},\ket{0,\mu},\ket{\mu,\mu}\}$ 
(with the first and second slot referring to sites $j$ and $j+1$, respectively), we obtain
\begin{align}
\hat{H}_{\rm p}^{(j,j+1)}\ket{0,0}&=0,\label{eq:Hp_to_00_OBC}\\
\hat{H}_{\rm p}^{(j,j+1)}\ket{\mu,0}&=2J^2(\ket{\mu,0}-\ket{0,\mu}),\label{eq:Hp_to_01_OBC}\\
\hat{H}_{\rm p}^{(j,j+1)}\ket{0,\mu}&=2J^2(\ket{0,\mu}-\ket{\mu,0}),\label{eq:Hp_to_10_OBC}\\
\hat{H}_{\rm p}^{(j,j+1)}\ket{\mu,\mu}&=0.\label{eq:Hp_to_11_OBC}
\end{align}
These matrix elements show that, within the one–magnon sector, the local parent Hamiltonian implements nearest-neighbor hopping of the magnon with amplitude $-2J^2$ between sites $j$ and $j+1$, while the vacuum and double-flipped configurations (i.e., states of the form $\ket{\mu,\mu}$) are annihilated.

For a magnon located in the bulk ($2\le j\le L-1$),
\begin{align}
\hat{H}_{\rm p}\ket{M_j^{\mu}}&=-2J^2(\ket{M^{\mu}_{j-1}}-2\ket{M^{\mu}_j}+\ket{M^{\mu}_{j+1}}).\label{eq:matrix_representation_Hp_one_magnon_sector_OBC_bulk}
\end{align}
At the open ends,
\begin{align}
\hat{H}_{\rm p}\ket{M^{\mu}_1}&=-2J^2(-\ket{M^{\mu}_1}+\ket{M^{\mu}_2}),\label{eq:matrix_representation_Hp_one_magnon_sector_OBC_left_edge}\\
\hat{H}_{\rm p}\ket{M^{\mu}_L}&=-2J^2(\ket{M^{\mu}_{L-1}}-\ket{M^{\mu}_L}).\label{eq:matrix_representation_Hp_one_magnon_sector_OBC_right_edge}
\end{align}
Collecting these relations into a matrix in the ordered basis $(\ket{M_1^{\mu}},\ket{M_2^{\mu}},\ldots,\ket{M_L^{\mu}})$, we obtain
\begin{align}
\hat H_{\rm p}^{\text{(1-mag)}}&\to -2J^2
    \begin{pmatrix}
        -1 & 1 & 0 & 0 & \cdots & 0\\
        1 & -2 & 1 & 0 & \cdots & 0\\
        0 & 1 & -2 & 1 & \cdots & 0\\
        \vdots & \vdots & \vdots & \vdots & \ddots & \vdots\\
        0 & 0 & 0 & 1 & -2 & 1\\
        0 & 0 & 0 & 0 & 1 & -1
    \end{pmatrix}\notag\\
&=-2J^2
    \begin{pmatrix}
        1 & 1 & 0 & 0 & \cdots & 0\\
        1 & 0 & 1 & 0 & \cdots & 0\\
        0 & 1 & 0  & 1 & \cdots & 0\\
        \vdots & \vdots & \vdots & \vdots & \ddots & \vdots\\
        0 & 0 & 0 & 1 & 0 & 1\\
        0 & 0 & 0 & 0 & 1 & 1
    \end{pmatrix} + 4J^2I\notag\\
    &\eqqcolon -2J^2A + 4J^2I,\label{eq:Matrix_representation_one-magnon_parent_Hamiltonian}
\end{align}
where $I$ is the $L\times L$ unit matrix. We next diagonalize the one-magnon Hamiltonian $\hat H_{\rm p}^{\text{(1-mag)}}$. Let ${\bm f}(p)=(f_1(p),f_2(p),\ldots,f_L(p))^T$ be an eigenvector of the matrix $A$ with eigenvalue $\lambda(p)$, i.e., $A{\bm f}(p)=\lambda(p){\bm f}(p)$. One finds
\begin{align}
f_j(p)&=\mathcal{N}_{p}\cos\left[\frac{\pi p}{L}\left(j-\frac{1}{2}\right)\right], \label{eq:expression_fj_OBC}\\
\lambda(p)&=2\cos\left(\frac{\pi p}{L}\right),\label{eq:expression_of_lambda_OBC}
\end{align}
with $p=0,1,\ldots,L-1$ and $\mathcal{N}_{p}$ being a normalization constant. Using these eigenmodes, we define a standing-wave creation operator
\begin{align}
\hat{F}^{\mu 0}_{\mathrm{OBC}}(p)&\coloneq\sum_{j=1}^L f_j(p)\hat{F}^{\mu 0}_j,\label{eq:definition_of_F10_OBC}
\end{align}
so that $\hat F^{\mu 0}_{\mathrm{OBC}}(p)\ket{\bm{0}}$ is a normalized one-magnon state. 
Direct calculations show that
\begin{align}
&[\hat{H}_{\rm p}^{\text{(1-mag)}},\hat{F}^{\mu 0}_{\mathrm{OBC}}(p)]\ket{\bm{0}}\notag \\
&=\left[-2J^2\lambda(p)+4J^2\right]\hat{F}^{\mu 0}_{\mathrm{OBC}}(p)\ket{\bm{0}}\notag\\
&=4J^2\left[1-\cos\left(\frac{\pi p}{L}\right)\right]\hat{F}^{\mu 0}_{\mathrm{OBC}}(p)\ket{\bm{0}}.\label{eq:RSGA_like_commutation_relation_OBC}
\end{align}
Therefore, the excitation energy is given by
\begin{align}
\mathcal{E}(p)&=4J^2\left[1-\cos\left(\frac{\pi p}{L}\right)\right],\label{eq:excitation_energy_OBC}
\end{align}
which is gapless and quadratic for $L\to\infty$:
\begin{align}
\mathcal{E}(p)&\sim 2J^2\left(\frac{\pi p}{L}\right)^2.\label{eq:thermodynamics_limit_excited_energy_OBC}
\end{align}
Equivalently, for fixed integer $p=O(1)$, $\mathcal{E}(p)\to 0$ as $L\to\infty$.

Although we have focused on the one-magnon sector, we can extend the results to the multimagnon sectors. To do this, we define
\begin{align}
\hat{F}^{\mu 0}_{\rm tot}\coloneq \sum_j\hat{F}_j^{\mu 0}.\label{eq:definition_of_F_tot}
\end{align}
Using the commutation relation $[\hat{H}_{\rm p},\hat{F}_{\rm tot}^{\mu 0}]=0$ and the fact that $\ket{\bm{0}}$ is a ground state of the SU($N$) ferromagnetic Heisenberg model, we can show that the following states are also ground states of the parent Hamiltonian:
\begin{align}
\ket{{\rm GS}_{\bm{m}}}\coloneq(\hat{F}^{10}_{\rm tot})^{m_1}(\hat{F}^{20}_{\rm tot})^{m_2}\cdots (\hat{F}^{N-1,0}_{\rm tot})^{m_{N-1}}\ket{\bm{0}},\label{eq:ground_state_of_SU(N)_Heisenberg_model}
\end{align}
where we define $\bm{m}\coloneq (m_1,m_2,\ldots,m_{N-1})$. Using the commutation relations $[\hat{H}_{\rm p}, \hat{F}^{\mu 0}_{\rm tot}]=0$ and $[\hat{F}^{\mu 0}_{\rm tot}, \hat{F}^{\nu 0}_{\rm PBC}(k)]=0$, we obtain
\begin{align}
[\hat{H}_{\rm p}, \hat{F}^{\mu 0}_{\rm PBC}(k)]\ket{{\rm GS}_{\bm{m}}}&=\mathcal{E}(k)\hat{F}^{\mu 0}_{\rm PBC}(k)\ket{{\rm GS}_{\bm{m}}}.\label{eq:RSGA-like_relation_multi_magnon_sector}
 \end{align}
This result implies that the state $\hat{F}^{\mu 0}_{\rm PBC}(k)\ket{{\rm GS}_{\bm{m}}}$ is an exact eigenstate of the parent Hamiltonian and has the same dispersion as in the one-magnon sector.
An analogous result holds under open boundary conditions by replacing $\hat{F}^{\mu 0}_{\rm PBC}(k)$ with $\hat{F}^{\mu 0}_{\rm OBC}(p)$ defined in Eq.~\eqref{eq:definition_of_F10_OBC}.

\subsection{Entanglement of the AQMBS states}\label{subsec:EE_of_AQMBS}
In this subsection, we quantify the entanglement of the gapless excited states under open boundary conditions introduced in Sec.~\ref{subsec:gapless_excitations}. Our strategy is to obtain the matrix product state (MPS) representation of the magnon excited states and evaluate the bond dimension $\chi$ of the MPS. It is well known that the von Neumann entanglement entropy (EE) is bounded by $\ln(\chi)$~\cite{schollwock2011density}. Here, we will show that the bond dimension is bounded polynomially in the system size, indicating that the EE of the magnon excited states obeys a subvolume law. These results are consistent with the identification of these states as AQMBS states in the SU($N$) Hubbard model.

First, we consider the MPS representation of the QMBS state (i.e., the ground state of the parent Hamiltonian), which is given by
\begin{align}
\ket{\tilde{S}_{\bm m}}&=\prod_{\mu=1}^{N-1} (\hat F^{\mu 0}_{\rm tot})^{m_\mu}|\bm{0}\rangle.\label{eq:QMBS_state_for_EE}
\end{align}

To obtain the MPS representation of $\ket{\tilde{S}_{\bm{m}}}$, we consider a matrix product operator (MPO) corresponding to the operator $\hat{F}^{\mu 0}_{\mathrm{OBC}}(p)$:
\begin{align}
\hat{F}^{\mu 0}_{\mathrm{OBC}}(p)&\coloneq\sum_{\bm{\ell},\bm{\ell}'}\hat{W}_0^{\mu}(\hat{W}_1^{\mu})^{\ell_1,\ell_1'}\cdots(\hat{W}_L^{\mu})^{\ell_L,\ell'_L}\hat{W}_{L+1}^{\mu}\ket{\bm{\ell}}\bra{\bm{\ell}'},\label{eq:MPO_representation_F_mu0_k}\\
\hat{W}_j^{\mu}&\coloneq
\begin{pmatrix}
\hat{I}_j & f_j(p)\hat{F}_j^{\mu 0} \\
0 & \hat{I}_j
\end{pmatrix}
,\quad \text{for }1\le j\le L,\label{eq:Matrix_form_of_Wj}\\
\hat{W}_0^{\mu}&\coloneq 
\begin{pmatrix}
1 & 0
\end{pmatrix}
,\quad \hat{W}_{L+1}^{\mu}\coloneq
\begin{pmatrix}
0 & 1
\end{pmatrix}^{\rm T}
,\label{eq:definition_of_edge_MPO}
\end{align}
where $\hat{W}_j^{\mu}$ is an operator-valued matrix, $\ell_j=0,1,\ldots,N-1$, and $\ket{\bm{\ell}}\coloneq \bigotimes_{j=1}^L\ket{\ell_j}_j$. Here, we used the technique of a finite state machine to obtain the above expression~\cite{Crosswhite2008finite,motruk2016density,paeckel2017automated}.
The MPO representation of $(\hat{F}^{\mu 0}_{\rm tot})^{m_{\mu}}$ is given by
\begin{align}
(\hat{F}^{\mu 0}_{\rm tot})^{m_{\mu}}&=\sum_{\bm{\ell},\bm{\ell}'}\hat{M}_0^{\mu}(\hat{M}_1^{\mu})^{\ell_1,\ell_1'}\cdots(\hat{M}_L^{\mu})^{\ell_L,\ell'_L}\hat{M}_{L+1}\ket{\bm{\ell}}\bra{\bm{\ell}'},\label{eq:MPO_representation_F_mu0_to_the_power_of_m_mu}
\end{align}
where $\hat{M}_j^{\mu}\;(1\le j\le L)$ is an operator-valued $(m_{\mu}+1)\times(m_{\mu}+1)$ matrix of the following form:
\begin{align}
\hat{M}_j^{\mu}&=
    \begin{pmatrix}
        \vspace{0.2em}\hat{I}_j & \hat F^{\mu 0}_j & 0 & 0 & \cdots & 0\\
        \vspace{0.2em}0 & \hat{I}_j & \hat F^{\mu 0}_j & 0 & \cdots & 0\\
        \vspace{0.2em}0 & 0 & \hat{I}_j & \hat F^{\mu 0}_j & \cdots & 0\\
        \vspace{0.2em}\vdots & \vdots & \vdots & \vdots & \ddots & \vdots\\
        \vspace{0.2em}0 & 0 & 0 & 0 & \hat{I}_j & \hat F^{\mu 0}_j\\
        \vspace{0.2em}0 & 0 & 0 & 0 & 0 & \hat{I}_j
    \end{pmatrix}
    ,\label{eq:MPO_representation_F_mu0_tot_m_mu}\\
    \hat{M}_0^{\mu}&=
    \begin{pmatrix}
    1 & 0 & \cdots & 0
    \end{pmatrix}
    ,\quad \hat{M}_{L+1}^{\mu}=
    \begin{pmatrix}
    0 & \cdots & 0 & 1
    \end{pmatrix}^{\rm T}.\label{eq:definition_of_M0_and_ML+1}
\end{align}
See, e.g., Sec. IV B of Ref.~\cite{Moudgalya2018}. From the above results, the bond dimension of the QMBS state $\ket{\tilde{S}_{\bm{m}}}$ becomes
\begin{align}
\chi_{\bm m}&=\prod_{\mu=1}^{N-1}(m_\mu+1),\label{eq:bond_dimension_of_QMBS}
\end{align}
where we used the fact that the bond dimension of the product of the MPOs is given by the product of the individual bond dimensions~\cite{schollwock2011density} and that $\ket{\bm{0}}$ is a direct product state. The bond dimension of the magnon excited state $\hat{F}^{\mu 0}_{\rm OBC}(p)\ket{\tilde{S}_{\bm{m}}}$ is at most $2\chi_{\bm{m}}$:
\begin{align}
\chi&\le 2\chi_{\bm m}=2\prod_{\mu=1}^{N-1}(m_\mu+1).\label{eq:bond_dimension_magnon_excited_state}
\end{align}
Therefore, the half-chain von Neumann EE $S_{\rm vN}$ obeys the inequality
\begin{align}
S_{\rm vN}\leq \ln\chi&\leq \ln 2 + \sum_{\mu=1}^{N-1}\ln(m_\mu+1).\label{eq:inequality_EE}
\end{align}
For fixed $N=O(1)$ and $m_\mu\leq L$, this gives
\begin{align}
    S_{\rm vN}\leq(N-1)\ln L + O(1),
\end{align}
which is a strictly subvolume law. Under the additional constraint $\sum_{\mu=1}^{N-1}m_{\mu}\le L$, the bound tightens further. These results establish that the gapless excitations built on $\ket{\tilde{S}_{\bm m}}$ are low-entanglement states and thus constitute AQMBS of the SU($N$) Hubbard model.

\section{Summary}\label{sec:summary}
We have extended the systematic construction of AQMBS states to SU($N$) Hubbard models with $N\geq 3$. Starting from the MLRSGA-$m$ and its symmetry-based formulation, we defined an enlarged overlap subspace $\mathcal{H}_P$ built from holons and flavor-selective doublons and showed that the tower of states $\{\ket{S_{\bm m}}\}$ forms exact QMBS within this setting.

Within $\mathcal{H}_P$, we constructed a parent Hamiltonian by projecting a local annihilator and proved that it reduces to the ferromagnetic SU($N$) Heisenberg model.
Thus, this establishes a new higher-symmetry family of parent Hamiltonians beyond the ubiquitous spin-1/2 case.

The low-energy sector of $\hat{H}_{\rm p}$ hosts gapless magnonlike modes. For periodic boundary conditions, one-magnon excitations created by $\hat{F}^{\mu0}_{\rm PBC}(k)$ have a dispersion $\mathcal{E}(k)=4J^2(1-\cos k)$, which is quadratic near $k=0$. For open boundary conditions, standing-wave excitations give $\mathcal{E}(p)=4J^2[1-\cos(\pi p/L)]$, which also vanishes in the limit $L\to\infty$ for fixed $p=O(1)$. These excitations are orthogonal to the towers of scar states and possess vanishing energy variance in the thermodynamic limit, thus satisfying the AQMBS criteria.

We further derived the upper bound on the entanglement of these excited states using MPS/MPO constructions. The scar state $\ket{S_{\bm m}}$ admits bond dimension $\chi_{\bm m}=\prod_{\mu=1}^{N-1}(m_{\mu}+1)$; acting with a single-mode MPO increases $\chi$ by at most a constant factor, implying $S_{\rm vN}\leq \ln\chi$. For fixed $N=O(1)$, the EE of the excited states exhibits sub-volume law scaling. This confirms that our gapless excitations are low-entanglement AQMBS above the SU($N$) scar tower.

Our results demonstrate the existence of AQMBS states in multiladder RSGA systems. Several directions for future work remain. Multiladder structures appear in pyramid scar states~\cite{Mark2020May,Omiya2023Fractionalization,Kunimi2024Oct,Odea2025Entanglement}. 

The construction of AQMBS states for these systems remains open, and our formulation would be useful for addressing this problem. The experimental observation of AQMBS in the SU($N$) Hubbard model is also an important direction. To this end, one would need to prepare an $\eta$-pairing state, since the AQMBS discussed here can be viewed as quasiparticle excitations above the $\eta$-pairing state. However, to the best of our knowledge, an experimental realization of $\eta$-pairing states has not yet been achieved even in the two-component Fermi-Hubbard model, making this a challenging task. We note that several proposals for preparing the $\eta$-pairing state in optical-lattice systems exist~\cite{Kantian2010,nakagawa2021eta}, and that the preparation of doublon states of bosonic atoms has been demonstrated experimentally~\cite{Su2023,honda2025observation}. Building on these techniques, the observation of AQMBS in SU($N$) Fermi-Hubbard systems may become feasible in the future.

\begin{acknowledgments}
The authors thank M. Nakagawa and A. Hokkyo for helpful discussions. This work was supported by JSPS KAKENHI Grant No.~JP25K00215 (M.K.) and JST ASPIRE No.~JPMJAP24C2 (M.K.).
\end{acknowledgments}

\section*{Data availability}
No data were created or analyzed in this article.

\appendix
\section{In the case of $N=3$}\label{sec:N3_case}
In this appendix, we investigate the specific structure of the SU(3) case and its connection to the general framework presented in the main text. For $N=3$, a representation-theoretic peculiarity allows for a more symmetric construction of the scar subspace. We show that this construction is equivalent to the $N=4$ case within the formalism developed in this paper.

As shown in Ref.~\cite{Nakagawa2024}, the additional QMBS states exist only for $N=3$. Let the three fermion flavors be $\{\alpha_1, \alpha_2, \alpha_3\}$. The ladder operators are defined by
\begin{align}
\hat{\eta}_{1,2}^\dagger &= \sum_j(-1)^j \hat{c}_{j,\alpha_1}^\dagger \hat{c}_{j,\alpha_2}^\dagger,\label{eq:definition_of_eta_1_2_app} \\
\hat{\eta}_{2,3}^\dagger &= \sum_j (-1)^j\hat{c}_{j,\alpha_2}^\dagger \hat{c}_{j,\alpha_3}^\dagger,\label{eq:definition_of_eta_2_3_app}\\
\hat{\eta}_{3,1}^\dagger &= \sum_j(-1)^j\hat{c}_{j,\alpha_3}^\dagger \hat{c}_{j,\alpha_1}^\dagger.\label{eq:definition_of_eta_1_3_app} 
\end{align}
The following state is also an eigenstate of the Hamiltonian (\ref{eq:definition_of_SU(N)_Hubbard_Hamiltonian}):
\begin{align}
\ket{\tilde{S}_{m_{1}, m_{2}, m_{3}}} = (\hat{\eta}_{1,2}^\dagger)^{m_{1}} (\hat{\eta}_{2,3}^\dagger)^{m_{2}} (\hat{\eta}_{3,1}^\dagger)^{m_{3}} \vac,\label{eq:additional_QMBS_N=3}
\end{align}
where $m_{1}$, $m_{2}$, and $m_{3}$ are nonnegative integers. These states are exact eigenstates (QMBS) of the SU(3) Hubbard Hamiltonian with energy $E = U \sum_{n=1}^3 m_{n}$, forming a tower of scar states.

The key difference from the general construction in the main text stems from the nature of the local two-particle states (doublons). For $N=3$, the three types of doublons, $\ket{\alpha_1\alpha_2}_j$, $\ket{\alpha_1\alpha_3}_j$, and $\ket{\alpha_2\alpha_3}_j$, transform as the conjugate fundamental representation $\bar{\mathbf{3}}$ of SU(3) \cite{Nakagawa2024}. This implies that all three doublon types are on an equal footing, and there is no need to single out a reference flavor as was done in the main text (e.g., $\alpha_1$).

The relevant Hilbert subspace $\mathcal{H}_P$ for the parent Hamiltonian is thus spanned by the holon state $\ket{0}_j = \vac_j$ and all three doublon states:
\begin{align}
\ket{1}_j &= \ket{\alpha_1\alpha_2}_j,\label{eq:definition_of_ket1_N=3} \\
\ket{2}_j &= \ket{\alpha_1\alpha_3}_j,\label{eq:definition_of_ket2_N=3} \\
\ket{3}_j &= \ket{\alpha_2\alpha_3}_j.\label{eq:definition_of_ket3_N=3}
\end{align}
This defines a local four-dimensional space $\{\ket{0}_j, \ket{1}_j, \ket{2}_j, \ket{3}_j\}$ at each site. The full subspace is given by $\mathcal{H}_P = \mathrm{Span}\{\bigotimes_j \ket{\ell_j}_j \mid \ell_j=0,1,2,3\}$.

Direct calculations show that the parent Hamiltonian for the SU(3) Hubbard model, when constructed in this symmetric basis, is identical to the SU(4) ferromagnetic Heisenberg model after an appropriate unitary transformation. 
This observation allows us to apply the results for the SU(4) ferromagnetic Heisenberg model to the SU(3) case without any modifications.
\bibliographystyle{apsrev4-2}
\bibliography{SU_N_AQMBS}
\end{document}